\documentclass[
    ,final            % use final for the camera ready runs
  ]
  {aipproc}

\layoutstyle{8x11single}

\begin{document}

\newcommand{\eg}{$E_{\rm \gamma}$}
\newcommand{\epo}{$E_{\rm peak}^{\rm obs}$}
\newcommand{\ep}{$E_{\rm peak}$}
\newcommand{\eiso}{$E_{\rm iso}$}
\newcommand{\liso}{$L_{\rm iso}$}

\title{Selection effects on GRB spectral-energy correlations}


\author{Lara Nava}{
  address={Universit\`a degli Studi dell'Insubria, via Valleggio 11, Como,
  Italy}
  ,altaddress={Osservatorio Astronomico di Brera, via E.Bianchi 46, Merate,
  Lecco, Italy}
}

\author{Giancarlo Ghirlanda}{
  address={Osservatorio Astronomico di Brera, via E.Bianchi 46, Merate,
  Lecco, Italy}
}

\author{Gabriele Ghisellini}{ address={Osservatorio Astronomico di Brera, via
    E.Bianchi 46, Merate, Lecco, Italy} }

\begin{abstract}
  Instrumental selection effects can act upon the estimates of the peak energy
  \epo, the fluence F and the peak flux P of GRBs.  If this were the case,
  then the correlations involving the corresponding rest frame quantities
  (i.e. \ep, \eiso\ and the peak luminosity \liso) would be questioned. We
  estimated, as a function of \epo, the minimum peak flux necessary to trigger
  a GRB and the minimum fluence a burst must have to determine the value of
  \epo\ by considering different instruments (BATSE, Swift, BeppoSAX). We find
  that the latter dominates over the former.  We then study the \epo-fluence
  (and flux) correlation in the observer plane. GRBs with redshift show well
  defined \epo-F and \epo-P correlations: in this planes the selection effects
  are present, but do not determine the found correlations. This is not true
  for Swift GRBs with redshift, for which the spectral analysis threshold does
  affect their distribution in the observer planes.  Extending the sample to
  GRBs without z, we still find a significant \epo-F correlation, although
  with a larger scatter than that defined by GRBs with redshift. We find that
  6\% are outliers of the Amati correlation. The \epo-P correlation of GRBs
  with or without redshift is the same and no outlier is found among bursts
  without redshift.
\end{abstract}

\maketitle

\section{Introduction}
Thanks to the discovery of the afterglow emission in long Gamma-Ray Bursts
(GRBs), redshift measurements of these sources became available. The knowledge
of GRB distances made possible to estimate their intrinsic properties and
several correlations between prompt and afterglow properties were found. One
of the most intriguing is the so called Ghirlanda correlation (\cite{gg04}),
linking the rest frame peak energy \ep\ to the collimated energy \eg\ emitted
during the prompt. The possibility to use this correlation for cosmological
purposes (\cite{gg04b}) makes it very appealing.  However, the possible
presence of selection effects, the lack of a physical interpretation, the
difficulty in calibrating the correlation and other problems mainly related to
the jet break time identification, raised objections about the validity of
this correlation and/or its application in cosmology. The lack of outliers
(except for GRB980425 and GRB031203, but see \cite{GG06}) and the small
dispersion of points around the best fit line are encouraging, but at this
stage the paucity of GRBs defining this correlation (27 GRBs) prevents us from
drawing any firm conclusion.

The paucity of GRBs with known \ep\ and \eg\ is partially due to the
difficulty to estimate the jet break times. They are expected in the afterglow
light curves at late times (about 1 day), when typically the light curve is
not well sampled. At late times, the observations are lacking and difficult
(e.g. for the presence of the host galaxy and/or a possible contamination from
the underlying supernova emission) and they are not sufficient to reveal the
presence of a break in the powerlaw decay behavior.  The existence of other
breaks and, more generally, the very complex temporal behavior discovered both
in the X-ray and optical light curves, makes the situation even more confused.
The different temporal decay of the X-ray and optical light curves raised the
question of a different origin for these two emissions.  \cite{GG07} suggested
that often the standard external shock model only accounts for the optical
emission, whereas the X-ray has another (internal) origin.  According to this
interpretation, \cite{gg07} caution about the identification of jet break
times, pointing out that they should be identified in the optical light
curves, at late times. According to the modeling of \cite{GG09}, these breaks
could also be chromatic.

When the jet break time is not available, we can only estimate the isotropic
equivalent energy \eiso, i.e. the energy emitted during the prompt by assuming
an isotropic emission geometry. The quantity \eiso\ still shows a correlation
with \ep\ (the so called Amati correlation, \cite{ama02}), in the form
\ep$\propto$\eiso$^{0.5}$. This correlation, found before the Ghirlanda one,
displays a larger scatter and a different slope. Assuming the validity of the
\ep-\eg\ correlation, the presence of this large scatter in the Amati
correlation can be easily explained, assuming that a wide range of jet opening
angles $\theta_j$ can correspond to bursts with same \eg\ (and therefore, for
the \ep-\eg\ correlation, same \ep). This range of $\theta_j$ values reflects
into a range of \eiso\ values, for the same given \ep.

Independently from its origin, the large dispersion of the Amati correlation
makes it unsuitable for the cosmological applications, unless systematics,
unknown, extra scatter terms are introduced.

Despite these problems, the Amati correlation appears robust. Although the
number of GRBs with known \ep\ and \eiso\ is reasonably large (83 objects in
\cite{n08}), only two of them do not obey the Amati correlation.  Note that
these two outliers are very peculiar GRBs (see \cite{GG06}).  Moreover, the
validity of the correlation has been extended to XRFs and now spans five
orders of magnitudes in \eiso, from XRFs to very bright GRBs. This behavior
calls for a theoretical explanation. Some works (\cite{butler07}, \cite{np05},
\cite{bp05}) suggest that its existence is due to the presence of selection
effects. To check this possibility it is necessary to perform a systematic and
accurate study of all the possible selection effects, to better understand if
and how they affect the correlation.

\begin{figure}
\includegraphics[height=.43\textheight]{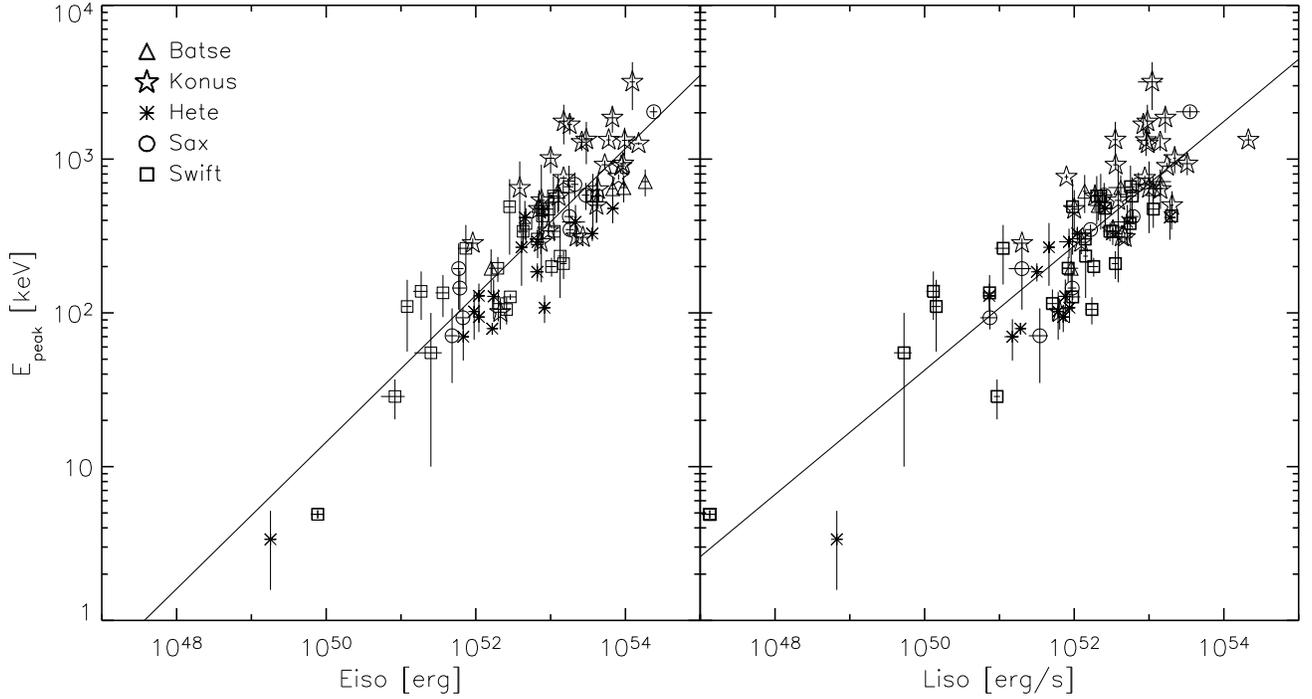}
\caption{\ep-\eiso\ and \ep-\liso\ correlations for 83 GRBs with measured
  redshift and spectral parameters. The slopes obtained from the fit with a
  power law model are respectively $0.48\pm 0.03$ and $0.40\pm 0.03$. Modeling
  the scatter distributions with a gaussian function we found respectively
  $\sigma=0.23$ and $\sigma=0.28$.}
\label{rf}
\end{figure}

Another correlation involving similar quantities is between \ep\ and \liso\ 
(Yonetoku correlation).  \cite{yone} found this correlation defining \liso\ as
the luminosity emitted at the peak of the light curve. In its original form,
this correlation appeared slightly tighter than the Amati one, but with a
similar slope. We will also study selection effects on this correlation. Fig.
\ref{rf} shows both the \ep-\eiso\ and the \ep-\liso\ correlations updated to
April 2008 (\cite{n08}).  The analysis performed on this sample (83 GRBs)
shows that the \ep-\liso\ correlation displays a larger scatter than the
\ep-\eiso\ correlation and therefore the latter seems to be more fundamental
(see \cite{n08} for details).

\newpage

\section{Instrumental selection effects}
\begin{figure}
\includegraphics[height=.43\textheight]{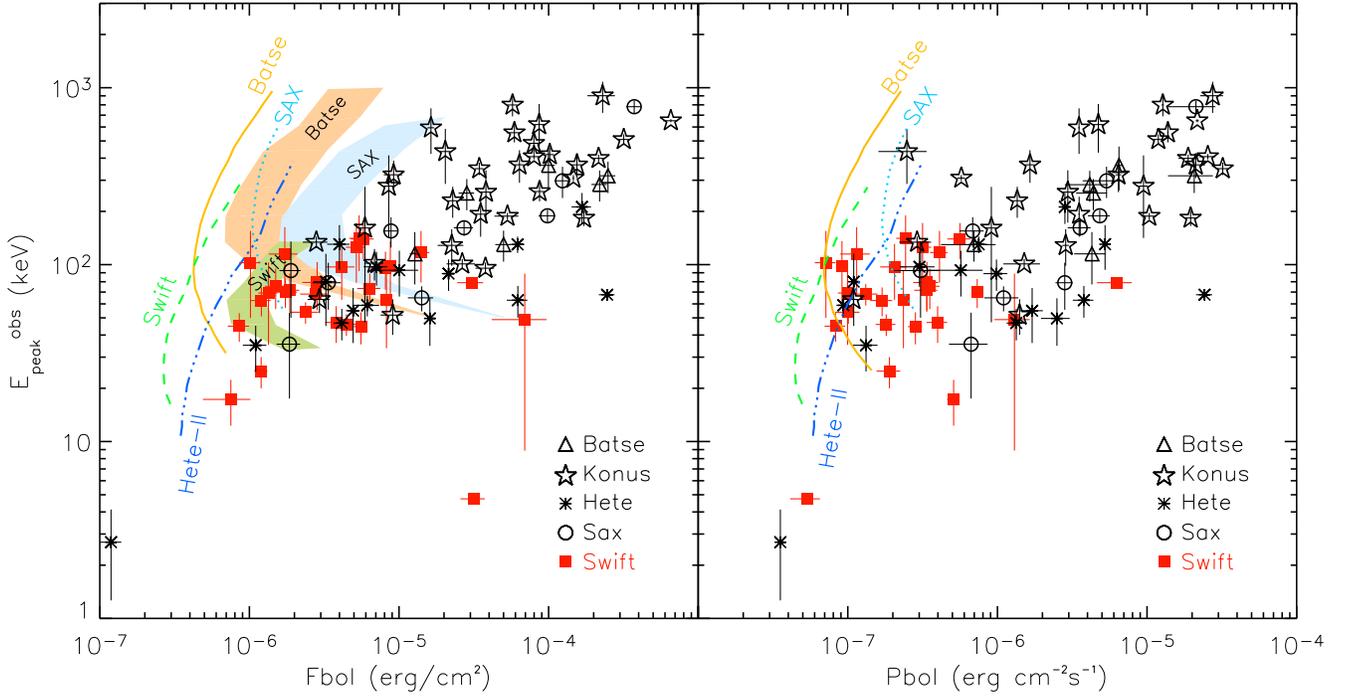}
\caption{Distributions of 83 GRBs with known redshifts in the \epo-Fluence and
  \epo-Peak flux planes. In both cases, they define a correlation in the
  observational planes. Shaded regions in the right panel show the ST (see
  text) for different instruments: for each instruments it is possible to
  perform a reliable spectral analysis only for bursts that lie on the right
  side of these regions.  In both the panels, the curves show the TT (see
  text) for different instruments: only bursts at the right side can be
  detected.}
\label{obs}
\end{figure}

We are interested in studying the rest frame correlations \ep-\eiso\ and
\ep-\liso.  However, instrumental selection effects act on the corresponding
observational quantities: \epo, Fluence (F) and Peak-Flux (P).  In Fig.
\ref{obs} we report the sample of 83 GRBs with known redshift plotted in the
observational planes \epo-F and \epo-P. Also in these planes, this sample of
GRBs defines a correlation.

Instrumental limits are obviously present. For example, very faint bursts can
not be detected by a given instrument if they are below its trigger threshold.
It is possible that the correlation is completely due to selection effects
that make accessible only a small part (a stripe) of the whole observational
plane. However, note that the region at the right side of the \epo-F (\epo-P)
correlation can not be affected by instrumental selection effects, since
bursts in this region would have an intermediate/low \ep\ (easily measurable
by instruments) and a high fluence (peak flux). The lack of bursts in this
region strongly suggests that bursts with these characteristics do not exist
or they are very few. On the left side of the two correlations we can face
with two different situations: the unavoidable selection effects acting in
these regions can i) lie far from the position of observed GRBs (in this case
the observed distribution of points is not determined by selection effects) or
ii) lie very close to points, indicating that they determine the shape of GRBs
distribution in this part of plane. To discriminate between these two cases,
we need to quantify the selection effects.

We identified the two most probable instrumental selection effects that can
affect the sample:
\begin{itemize}
\item the trigger threshold (TT): it is very complex to quantify what
  characteristic a burst must have to be detected by a given instrument. To a
  first approximation, we can affirm that it must have a minimum photon peak
  flux.  This translate into a minimum energy flux, whose value depends on the
  spectral properties of the burst (especially \epo). For several instrument
  \cite{band03} derive the curves of minimum peak flux as a function of \epo.
  We plot these curves in the \epo-P plane, after accounting for the
  bolometric correction.  Considering a typical conversion factor between
  fluence and peak flux of $\sim 6$ (\cite{gg08}), we plot these curves also
  in the \epo-Fluence plane.
\item the spectral analysis threshold (ST): we need a minimum fluence to
  perform a reliable spectral analysis and determine \epo\ and the spectral
  shape. \cite{gg08} consider this problem by accounting for the detector
  response and the typical background of each instrument. For each \epo\ they
  determine this limiting fluence and obtain a curve in the \epo-F plane. Its
  position depends on the burst duration. In Fig.~\ref{obs} (left panel) we
  show the ST as a shaded region: the left (right) edge represents the curve
  obtained for bursts lasting 5 (20) seconds.
\end{itemize}

Comparing the position of points in Fig.~\ref{obs} detected by a given
instrument with the corresponding threshold curves we can draw several
conclusions. In the \epo-F plane the ST represents a more important selection
effect compared to the TT. Swift bursts (filled squares) are strongly affected
by the corresponding ST limiting curves. Note that for Swift bursts we mean
bursts whose \epo\ has been determined from the modeling of the BAT data.
However, this instrument can only measure \epo\ within the range 15-150 keV.
Consequently, the Swift sample is distributed in a very narrow range in \epo,
smaller than the scatter of the \epo-F correlation as defined by all the
pre-Swift bursts. For this reason, we caution about the use of Swift bursts to
test the existence of this correlation.

Our results also show that the pre-Swift GRB sample, containing a fraction of
bursts detected by BeppoSAX and BATSE, is not affected by the corresponding
limiting curves. This is particularly evident for high \epo, where the
limiting curves lie far from the points. A similar conclusion can be drawn for
the \epo-P correlation (right panel). Points are slightly affected by the
corresponding TT in the region at low peak flux, but the behavior according
to which higher peak fluxes correspond to higher \epo\ is not due to the
threshold.

In both planes however, the lack of bursts (with known redshift) with
intermediate/large \epo\ and small fluence or peak flux (i.e.  between the
present sample of GRBs and the limiting curves) could be due to the additional
presence of a still-not-understood selection effect, for example linked to the
redshift determination. The first step to investigate this possibility is to
search if there exist GRBs which populate the region on the left-hand side of
the \epo-F and \epo-P correlations. This issue can be solved by including, in
both plots, all the bursts with measured spectral properties, without the
request of knowing their redshifts.
                                                    
\section{GRBs without redshift measurement}
In this section we study the very same issue tackled in the previous one, but
considering all bursts found in literature with measured \epo\ and spectral
properties. For this reasons, we considered Hete-II bursts (\cite{saka05}),
Swift bursts (\cite{butler07}) and the sample of bright BATSE bursts
(\cite{kane06}). Moreover, we collected the preliminary results arising from
the analysis of the Konus-Wind spectra as reported in the GCN circulars (29
GRBs). To extend the sample of BATSE bursts to lower fluences we performed the
spectral analysis of 100 BATSE bursts down to $10^{-6}$erg/cm$^2$ (see
\cite{n08} for details).

\begin{figure}
\includegraphics[height=.43\textheight]{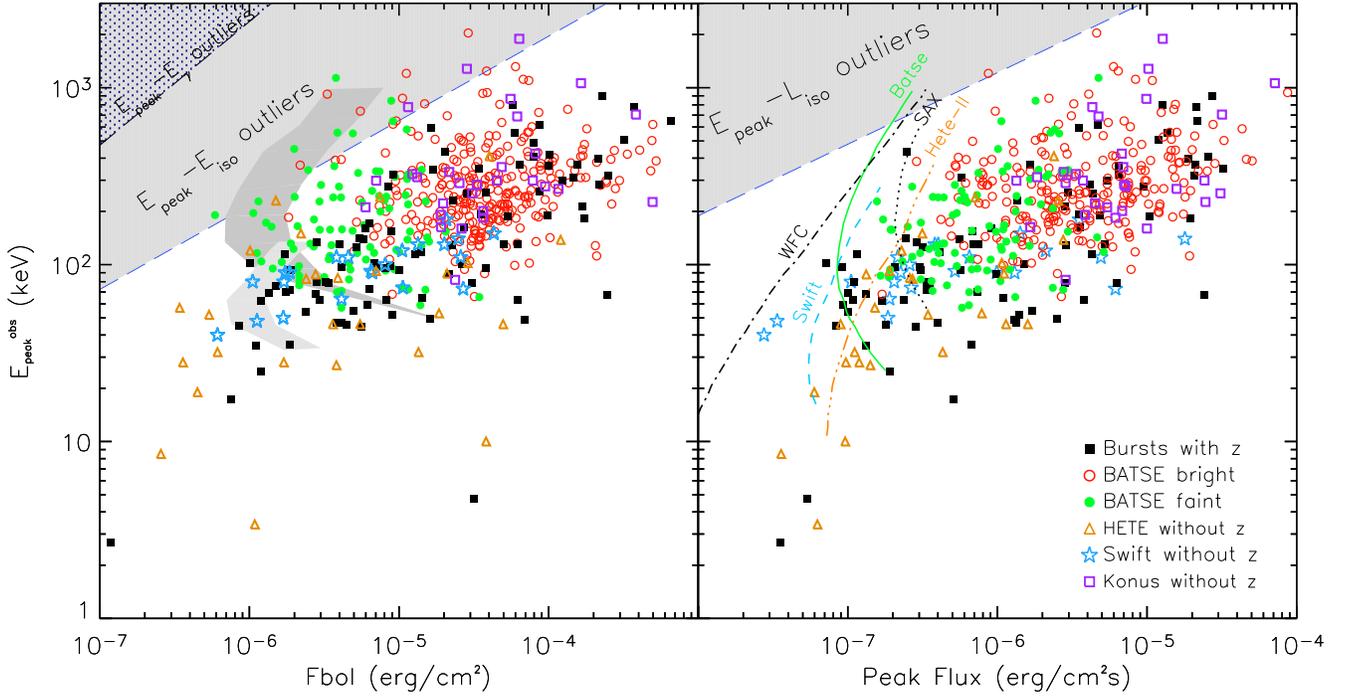}
\caption{Left panel: bursts with (filled squares) and without (other symbols)
  known redshift in the \epo-Fluence plane. Shaded curves represent the ST
  curves for BATSE and Swift/BAT instruments (see Fig.~\ref{obs}). Shaded
  regions in the top left corner show which bursts can not be consistent with
  the 3$\sigma$ scatter of the Amati relation defined in the rest frame plane
  by GRBs with redshift. Right panel: \epo-Peak flux plane. Curves show for
  several instrument the TT, i.e. the minimum peak flux required to detect the
  burst.  In this case only one burst fall in the region of outliers.}
\label{obstutti}
\end{figure}

Fig.~\ref{obstutti} (left panel) reports the distribution of all these bursts
in the \epo-F plane. Different symbols refers to different instruments. We
note that the entire sample of GRBs without redshift seems to occupy a larger
region with respect to GRBs with redshift (filled squares) studied in the
previous section. In particular, they enlarge the distribution toward the ST
curves. However, moving towards lower fluence values their density decreases
(see \cite{n08} for more details). In other words, points do not uniformly
fill all the accessible plane, but are concentrated along a strip whose slope
and position is not determined by the considered selection effects.
Therefore, the existence of the Amati correlation must have a physical origin.
However, the dispersion of the whole sample of GRBs is larger then that
defined only by GRBs with redshift. This suggests that, if we were able to
measure the distance of all GRBs in Fig.~\ref{obstutti}, we should found an
Amati correlation with different properties, such as a larger scatter and,
likely, a different slope. The grey region in the top left corner of the
\epo-F plane (see Fig.~\ref{obstutti}) shows the region where bursts are
outliers of the Amati correlation (as defined by the sample of 83 GRBs with
redshift). This means that GRBs that lie in this region can not be consistent
with the correlation for any redshift.  For 'consistent' we mean the burst
must fall within the 3$\sigma$ scatter of the correlation. We found that the
6\% of the sample is not consistent with the Amati correlation.

The right panel of Fig.~\ref{obstutti} refers to the \epo-Peak flux plane.
Also in this case it is evident the presence of a correlation, i.e. of a
distribution of points that preferentially lie along a strip. This behavior
cannot be explained in terms of selection effects, since the shape and the
position of the TT curves cannot account for the observed distribution of
points. The difference with respect to the case of the \epo-F correlation is
that, here, the distribution of points without redshift seems to better agree
with that of GRBs with redshift. This suggests that in the future, when a
larger sample of GRBs with redshift will be available, the properties of the
Yonetoku correlation (slope and scatter) will resemble those of the present
correlation.  Also in this plane we have estimated the regions of outliers and
we found only one burst which cannot be consistent with the correlation.

\section{Conclusions}
To study the role of possible instrumental selection effects on the Amati and
Yonetoku relation, we have considered the distribution of GRBs with and
without known redshift in the observational \epo-Fluence and \epo-Peak flux
planes. Following the analysis performed by \cite{gg08}, we refer to two
different instrumental biases: the trigger threshold (TT, the minimum peak
flux required to trigger a burst) and the spectral analysis threshold (ST, the
minimum fluence needed to constrain the GRB spectral properties). These curves
depend on \epo\ and define what part of the observational planes is
accessible: only bursts at the right side of both the curves can be plotted in
these planes, the other bursts have no sufficient flux to be detected or no
sufficient fluence to recover \epo\ and the spectral shape from the spectral
analysis. We note that in both the observational planes GRBs are not spread in
the region free from instrumental selection effects, but define a correlation.
Note that the shape of this concentration of points is not determined by the
considered thresholds. Their only effect is to cut the part of the
correlations corresponding to the smallest \epo\ and Fuences/Peak fluxes.
From the comparison between bursts with and without redshift we can conclude
that there exists an \ep-\eiso\ correlation not determined by selection
effects, even if its slope and scatter may be different from what we know now.
On the other hand, we suggest that, once a large number of bursts with
redshift will be available, the Yonetoku correlation will preserve more or
less the present scatter and slope.

Another hint in favor of the relevance of the \ep-\liso\ correlation comes
from short bursts. A detailed analysis of a large sample of short bursts
performed by \cite{gg09} shows that they have a similar \epo\ and Peak flux of
long GRBs and, indeed, they populate the same region in the \epo-Peak flux
plane. This suggests they can be consistent with the same \ep-\liso\ 
correlation defined by long GRBs (this is also confirmed by the few short
bursts for which we know the redshift).  Short GRBs, instead, are inconsistent
with the distribution of long GRBs in the \epo-Fluence plane, since, for the
very same \epo\ they have a smaller fluence. This implies that the majority of
short GRBs are outliers of the \ep-\eiso\ correlation defined by long bursts.

\end{document}